\documentclass[assymb,11pt]{article}
\usepackage{amssymb}
\usepackage[pdftex,colorlinks=true,urlcolor=black,filecolor=black,linkcolor=black,
            pdftitle={The Relative Motion of Membranes.},
            pdfauthor={Mark D. Roberts},
            pdfsubject={relativity, quantum theory},
            pdfkeywords={membrane separation, quantization},
            pagebackref,pdfpagemode=None,bookmarksopen=true]{hyperref}
\newcommand{\bc}{\begin{center}}
\newcommand{\ec}{\end{center}}
\newcommand{\bi}{\begin{itemize}}     
\newcommand{\ei}{\end{itemize}}
\newcommand{\bd}{\begin{description}} 
\newcommand{\ed}{\end{description}}
\newcommand{\bn}{\begin{enumerate}}   
\newcommand{\en}{\end{enumerate}}
\newcommand{\be}{\begin{equation}}
\newcommand{\ee}{\end{equation}}
\newcommand{\ber}{\begin{eqnarray}}
\newcommand{\ear}{\end{eqnarray}}
\newcommand{\ba}{\begin{array}}
\newcommand{\ea}{\end{array}}
\newcommand{\bv}{\begin{verbatim}}
\newcommand{\ev}{\end{verbatim}}

\newcommand{\al}{\alpha}

\newcommand{\bt}{\beta}
\newcommand{\bx}{\square}
\newcommand{\ch}{\chi}
\newcommand{\de}{\delta}
\newcommand{\De}{\Delta}

\newcommand{\dt}{\rm{d}\tau}
\newcommand{\Dt}{\frac{\rm{D}}{\rm{d}\tau}}
\newcommand{\el}{\ell}

\newcommand{\fr}{\frac}

\newcommand{\ga}{\gamma}

\newcommand{\La}{\Lambda}

\newcommand{\Lg}{{\cal L}}
\newcommand{\n}{\nonumber\\}

\newcommand{\na}{\nabla}

\newcommand{\ps}{\psi}

\newcommand{\Si}{\Sigma}

\newcommand{\sq}{\sqrt}

\newcommand{\ta}{\tau}

\newcommand{\p}{\partial}
\begin{document}
\title{The Relative Motion of Membranes.}
\author{
\href{http://www.violinist.com/directory/bio.cfm?member=robemark}
{Mark D. Roberts},\\
54 Grantley Avenue,  Wonersh Park,  GU5 0QN,  UK\\
mdr@ihes.fr
}
\date{$16^{th}$ of March 2010}
\maketitle
\begin{abstract}
The relative classical motion of membranes is governed by the equation
$(w^{\al~c}_{~\bt~c}r^{\bt a})_a=R^\al_{~\de\ga\bt}r^\bt x^{\de a}p^\ga_a$,
where $w$ is the hessian.
This is a generalization of the geodesic deviation equation and can be derived from
the lagrangian $p\cdot \dot{r}$.
Quantum mechanically the picture is less clear.
Some quantizations of the classical equations are attempted so that
the question as to whether the Universe started with a quantum fluctuation
can be addressed.
\end{abstract}
\section{Introduction}
\label{intro}
The geodesic deviation equation can be thought of as a way of expressing the Ricci identity.
It can be much generalized to equations describing relative motion of many sorts of system;
the easiest way of doing this is by the lagrange method,
using a lagrangian of the form $p\cdot \dot{r}$,
where $p$ is the momentum of a given system and $r$ is a seperation.
Here attention is restricted to the given system being a membrane
so that questions in contemporary membrane cosmology \cite{KOST} can be approached.
The geodesic equations are sometimes thought of as a consequence of field equations \cite{EIH}.
So far the geodesic deviation equations appear to be independent of any set of field equations,
and the same is true of its generalizations.
This could be thought of as a good thing,
as it means that any results involving them are indepedent of specific field equations,
or as a bad thing as it does not allow choices to be made as to which field equations are best.

In \S\ref{cca} the general lagrangian theory allowing relative motion equations
to be derived is presented.
The main feature of this theory is that the variables
are cross-conjugate rather than self-conjugate:
this mean that the momentum associated with a given system $p$
is conjugate to the separation $r$
and that the momentum associated with the extension $P$ is conjugate to $x$.
This causes all sorts of technical problems,
because it does not allow the separation part to be added on linearly and approximated.
The extended theory is a gauge theory,
as has been described for the point particle \cite{mdr24}.
Two systems in particular are used to illustrate cross-conjugate lagrangians,
the point particle and the membrane,
strings having been looked at previously \cite{mdr26}.
In \S\ref{maxsym} the resulting relative motion equations
are calculated in the case of maximal symmetry.
Here the relative motion of membranes is reduced right down so that it is similar
to that of point particles;
however membranes can have their own dynamical degrees of freedom \cite{GV1}.
In \S\ref{quant} attempts are made to quantize the theory.
There seems to be five approaches to this.
The first is if one thinks as the relative motion equations
as being embedded in some general set of field equations,
quantize these field equations,  this is not looked at here.
The second is that the relative motion lagrangian is an example of a gauge theory.
This allows it to be quantized by formal methods,
this has been done for the point particle \cite{mdr24}.
The problem with this is that, although the Klein-Gordon equation can be recovered,
the resulting wavefunctions do not seem to be separable into $x$ and $r$ parts;
also it assumes that one knows how to quantize the given system.
This approach is not looked at because of lack of wavefunction separability.
The third is to assume that methods used to quantize the given system
can be applied to the extended system,
this has been done for the point particle \cite{BK}.
Here a naive quantization of the membrane is applied to the extended system.
The fourth is to quantize membrane fluctuations \cite{GV2}.
The fifth is to apply assorted substitutions
which give back plausible looking classical equations,
this is done here to try to quantize the extended part of the system.
\S\ref{conc} is the conclusion.
Things not looked at include:
firstly,
any derivation of relative motion equations by second variation of standard lagrangians,
this has been done for strings \cite{mdr26},
secondly,
anything to do with spin,  fermions,  torsion, vector fields,  or fluids,
geodesic deviation has been generalized to include spin \cite{mohseni},
thirdly,
any study of gravitational waves \cite{NSV},
fourthly,
any detailed investigation of the relationship between relative motion equations
and field equations,
fifthly,
any application of the equations to specific configurations except maximal symmetry.
For a lagrangian theory $\Lg$,  with action $I$,  the momentum and hessian are:
\be
p^\al\equiv\fr{\de I}{\de\dot{q}^\al}=\fr{\p\Lg}{\p\dot{q}^\al},~~~~~~~
w^{\al\bt}\equiv\fr{\de^2 I}{\de\dot{q}^\al\de\dot{q}^\bt}
               =\fr{\p^2\Lg}{\p\dot{q}^\al\p\dot{q}^\bt},
\label{1.1.1}
\ee
where $q=x$ if the lagrangian is that of an extended object.
Small letters $p,w$ are used for the given system,
and capitals $P,W$ for the system extended to a cross-conjugate deviating system.
Roman p used for the p-brane.
Greek letters are used for spacetime indices,  latin letters for internal indices.
Analogs and generalizations of the geodesic deviation equation are here called relative motion
equations,  rather than deviation equations.
Lower case "lagrangian" and "hessian" throughout.
All other notation is as in Hawking and Ellis (1973) \cite{HE}.
\section{Cross-Conjugate Actions.}
\label{cca}
The lagrangian is generalized from that of a given system $\Lg$
to the relative motion lagrangian $\Lg_r$,  the action is taken as
\be
S=\int^{\ta_2}_{\ta_1}\dt\Lg_r,~~~
\Lg_r=a_1\Lg+a_2\dot{r}\cdot p.
\label{2.1.1}
\ee
$r$ is a separation vector,  $a_1$ and $a_2$ and constants,
typically $a_1=0$ as the equation of motion $\dot{p}=0$
follows from the second part of (\ref{2.1.1}),  and $a_2=1$.
The lagrangian $\Lg$ has momentum from which a cross-product has to be formed with $\dot{r}$,
typically this means that the momentum has one spacetime index so that
$\dot{r}\cdot p=\dot{r}^\al p_\al$.
The integral in (\ref{2.1.1}) takes that form for the point particle,
however for membranes it is over all the internal indices,
and for fields it is usually over some universal time t;
similarly the dot on $r$ is proper time for the point particle,
all the internal indices for a membrane,  i.e. $\dot{r}\rightarrow\p_ar$,
and for fields it usually represents differentiation with respect to some universal time.

For $\Lg_r=\Lg_r(\dot{x},\dot{r})$ the Ricci identity is
\be
\De \dot{r}^\al=\Dt\De r^\al+R^\al_{.\bt\ga\de}r^\bt\de x^\ga\dot{x}^\de.
\label{2.1.2}
\ee
Using (\ref{2.1.2}) $\de x$ and $\De r$ variations of (\ref{2.1.1}) give
\be
-\Dt \fr{\p \Lg_r}{\p \dot{x}^\al}
+R_{\ga\bt\al\de}r^\bt\dot{x}^\de\fr{\p \Lg_r}{\p \dot{r}^\ga}=0,~~~
-\Dt\fr{\p\Lg_r}{\p\dot{r}^\al}=0,
\label{2.1.3}
\ee
respectively.
For the specific form of the Lagrangian in (\ref{2.1.1}),
substituting (\ref{1.1.1}) and (\ref{2.1.1}) into (\ref{2.1.3})
gives the momenta and hessian
\ber
P^\al_r&=&\fr{\p\Lg_r}{\p\dot{r}^\al}
        =a_2\fr{\p\Lg}{\p\dot{x}^\al}=a_2p^\al,\n
P^\al_x&=&\fr{\p\Lg_r}{\p\dot{x}^\al}
        =a_1\fr{\p\Lg}{\p\dot{x}^\al}+a_2\fr{\p^2\Lg}{\p\dot{x}^\al\p\dot{x}^\bt}\dot{r}^\bt
        =a_1p^\al+a_2w^\al_{.\bt}\dot{r}^\bt,\n
W^{\al\bt}_{rr}&=&0,\n
W^{\al\bt}_{ru}&=&W^{\al\bt}_{ur}=a_2w^{\al\bt},\n
W^{\al\bt}_{uu}&=&a_1w^{\al\bt}
+a_2\dot{r}^\ga\fr{\p^3\Lg}{\p\dot{x}^\al\p\dot{x}^\bt\p\dot{x}^\ga},
\label{2.1.4}
\ear
which allows the general form of the relative motion equations (\ref{2.1.3})
to be expressed in term of the momentum
\be
\dot{p}^\al=0,~~~
\dot{P}^\al_x=a_2R^\al_{.\de\ga\bt}r^\bt\dot{x}^\de p^\ga,
\label{2.1.5}
\ee
where $P^\al_x$ is given by (\ref{2.1.4}) and so has the hessian of the original system in it.
In general metric variation of (\ref{2.1.1})
gives stresses which do not have an immediate interpretation.

For the point particle the lagrangian can be taken as
\be
\Lg=-m\el,~~~\el\equiv\sq{-\dot{x}^2}.
\label{2.2.1}
\ee
From (\ref{1.1.1}),  (\ref{2.2.1}) yields momentum and hessian
\be
p^\al=\fr{m\dot{x}^\al}{\el},~~~
w^{\al\bt}=\fr{m}{\el}h^{\al\bt},
\label{2.2.2}
\ee
where $h^{\al\bt}$ is the specific projection tensor from general relativity
\be
h^{\al\bt}\equiv g^{\al\bt}-\fr{\dot{x}^\al\dot{x}^\bt}{\dot{x}^2}.
\label{2.2.3}
\ee
The momentum constraint is
\be
p^2+m^2=0.
\label{2.2.a}
\ee
Substituting (\ref{2.2.2}) and (\ref{2.2.3}) into (\ref{2.1.4}) gives the extended momentum
\be
P^\al_x=a_1p^\al+a_2\fr{m}{\el}\dot{r}^\bt h^\al_{.\bt}
       =\left(a_1+a_2\fr{\dot{r}\cdot\dot{x}}{\el^2}\right)p^\al+a_2\fr{m}{\el}\dot{r}^\al,
\label{2.2.4}
\ee
which obeys the relative motion equation (\ref{2.1.5}).
The extended momentum constraint is
\be
p_\al P^\al_x=-a_1m^2.
\label{2.2.b}
\ee
Changing notation $(P_x,p,a_1,a_2,\al,\de,\ga,\bt)\rightarrow(\Pi,P,0,1,\mu,\al,\bt,\ga)$
gives the geodesic deviation equation (\ref{2.1.5}) in the form previously studied by
the author (\cite{mdr26}).

The $p$-dimensional membrane action can be taken in the form
\be
S=k\int_{\cal M}d^{p+1}\xi\sq{-\ga},
\label{2.3.1}
\ee
\cite{mdr33},  where $\ga$ is the determinant of the internal metric
$\ga^{ab}=g_{\mu\nu}x^{\mu a}x^{\nu b}$,  and the internal indices are $a,b,\ldots$.
The momentum and hessian are
\be
p^{\mu a}=\fr{\p\Lg}{\p x^{\mu a}}=-k\sq{-\ga}x^{\mu a},~~~
w^{\mu\nu ab}=\fr{\p^2\Lg}{\p x^{\nu b}\p x^{\mu a}}
             =-k\sq{-\ga}(g^{\mu\nu}\ga^{ab}-x^{\mu a}x^{\nu b}),
\label{2.3.2}
\ee
The momentum constraints are
\be
p^{\mu a}p^b_\mu+k^2\ga\ga^{ab}=0,
\label{2.3.c}
\ee
producing $\Si(p+1)$ constraints,
fixing $a=\ta$ gives the usual $(p+1)$ first class constraints.
When $a=b$ (\ref{2.3.c}) gives $p^2+({\rm p}+1)k\ga=0$,
which reduces to (\ref{2.2.a}) for the point particle.
The relative motion Lagrangian is
\be
\Lg_r=p^{\mu a}\p_ar_{\mu},
\label{2.3.3}
\ee
the extended momentum is
\be
P^{\al a}_x=w^{\al~c}_{.\bt~.c}\p^ar^\bt.
\label{2.3.4}
\ee
The extended momentum constraint is
\be
p_{\mu a}P^{\mu a}_x+k{\rm p}\sq{-\ga}r^{\bt b}p^a_\bt=0.
\label{2.3.b}
\ee
Define the first fundamental form $\ch^{\mu\nu}\equiv x^{\mu a}x^\nu_{.a}$,
and the projection tensor $h^{\mu\nu}\equiv g^{\mu\nu}-\ch^{\mu\nu}$,
then the hessian contracted over internal indices takes the form
\be
w^{\mu\nu c}_{.~.~.c}=-k\sq{-\ga}(h^{\mu\nu}+pg^{\mu\nu}),
\label{2.3.a}
\ee
and this is proportional to the projection tensor only in the case of the point particle.
The equations of relative motion are
\be
\p_ap^{\mu a}=0,~~~
\p_aP^{\mu a}_x=R^\mu_{.\de\ga\bt}r^\bt p^{\ga a}\p_ax^\de,
\label{2.3.5}
\ee
with the internal indice ''a'' summed up on each side,
so that the Riemann tensor occurs only once.
The string deviation equation \cite{mdr26},  is an example of (\ref{2.3.5}).
\section{Maximally Symmetric Case.}
\label{maxsym}
For maximal symmetry
\be
R_{\al\bt\ga\de}=\fr{\La}{3}(g_{\al\ga}g_{\bt\de}-g_{\al\de}g_{\bt\ga}).
\label{3.1}
\ee

Taking only the $a_2$ term, using the definition of the point particle momentum (\ref{2.2.2}),
and dividing by $a_2$ gives the geodesic deviation equation
\be
\fr{\rm d}{{\rm d}\ta}P^\al_x
=m\fr{\rm d}{{\rm d}\ta}\left(\fr{\dot{r}^\bt h^\al_\bt}{\el}\right)
=-\fr{m\La}{3}\el h^\al_\bt r^\bt.
\label{3.2}
\ee
with projection tensor given by (\ref{2.2.3}).
In the case $\dot{x}\cdot\dot{r}=\dot{x}\cdot r=0$,
equivalently $r^\al=h^\al_\bt r^\bt,\dot{r}^\al=h^\al_\bt\dot{r}^\bt$,
the relative motion of the two particles is orthogonal to their direction of propagation,
also taking $\el=1$ (\ref{3.2}) becomes
\be
\ddot{r}^\al=-\fr{\La}{3}r^\al,
\label{3.3}
\ee
which has solution
\be
r^\al=\hat{r}^\al\left[C_+\exp\left(+\sq{-\fr{\La}{3}}\ta\right)
                  +C_-\exp\left(-\sq{-\fr{\La}{3}}\ta\right)\right].
\label{3.4}
\ee
For negative cosmological constant,  i.e.$\La<0$,
and $C_-=0$ the particles slowly approach one another;
for $\La>0$ they oscillate.

For the membrane (\ref{3.2}) generalizes to
\be
\na_a\left(w^{\al~c}_{.\bt.c}r^{\bt a}\right)=-\fr{\La}{3}w^{\al~c}_{.\bt.~c}r^\bt,
\label{3.5}
\ee
the hessian taking place of the projection tensor.
Again assuming orthogonal conditions $\ch^\mu_\bt r^\bt=\ch^\mu_\bt r^{\bt a}=0$.
Now assume $\sq{-\ga}=1$ and divide by $(p+1)$ to arrive at
\be
r^{\al a}_{.~.a}=-\fr{\La}{3}r^\al
\label{3.6}
\ee
which governs orthogonal motion between membranes.
If the internal spatial derivatives are negligible solutions like (\ref{3.4}) can be found.
\section{Quantum Theory.}
\label{quant}
In the already quantized approach one assumes that the prescription used to quantize the given
system $\Lg$ can also apply to the extended relative motion system $\Lg_r$.
For the point particle this prescription is
\be
p_\al\rightarrow-i\hbar\na_\al.
\label{4.1.1}
\ee
The constraint (\ref{2.2.a}) gives the Klein-Gordon equation,
now there is in addition the extended constraint (\ref{2.2.b}).
It is not clear what substitutions should be applied to $P^\al_x$,
so it is left as a classical object.
Taking $a_1=0,a_2=1$ and then applying the operator substitutions (\ref{4.1.1}) to (\ref{2.2.b})
and ordering $P_x$ before $p$ gives
\be
P^\al_x\ps_\al=0,
\label{4.1.b}
\ee
ordering $p$ before $P_x$ gives the wrong classical limit.
Substituting $\dot{x}$ everywhere with the momentum by using (\ref{2.2.2}),
the geodesic deviation equation (\ref{2.1.5}) and (\ref{2.2.4}) becomes
\be
\fr{\rm d}{{\rm d}\ta}\left[\fr{m}{\el}\dot{r}_\bt
                            \left(g^{\al\bt}+\fr{p^\al p^\bt}{m^2}\right)\right]
=\fr{\el}{m}R^\al_{.\de\ga\bt}r^\bt p^\de p^\ga.
\label{4.1.3}
\ee
Replacing $p$ via the operator substituting (\ref{4.1.1}),  and choosing the differential
operator to operate twice on the wavefunction,  rather than separately on wavefunctions,
also that the wavefunction is inside the square bracket on the rhs of (\ref{4.1.3}),
and also the $\el$'s,  the Riemann tensor and $\dot{r}$
are taken to the left of the differential operators gives
\be
\fr{\rm d}{{\rm d}\ta}\left[\fr{m}{\el}\dot{r}_\bt
                            \left(g^{\al\bt}-\fr{\hbar^2}{m^2}\na^\al\na^\bt\right)\ps\right]
=-\fr{\hbar^2\el}{m}R^\al_{.\de\ga\bt}r^\bt\na^\de\na^\ga\ps.
\label{4.1.4}
\ee
Using the principle function $S$,  with $\ps=\exp(iS/\hbar)$,
(\ref{4.1.3}) is recovered in the classical limit $\hbar\rightarrow0$
if in addition $S,_\al=p_\al$ and $(\ln\ps)_\ta$ is small.
That the principle function only recovers cases where the momentum $p$ is a gradient
vector also happens for the classical limit of the Klein-Gordon equation.
The extended momenta $P_x$ has to be treated as a separate object
in the extended constraint (\ref{4.1.b}),
otherwise it gives the term in the square bracket ''[]'' in (\ref{4.1.4}) vanishes.

For the membrane $p^{\al a}\rightarrow-i\hbar\p^a\na^\al$
introduces derivatives of higher order,
however the substitution
\be
p^{\al a}\rightarrow-i\hbar v^a\na^\al,~~~
v^av^b=\ga\ga^{ab},
\label{4.6.1}
\ee
allows derivatives of the wavefunction to be of the same order as for the point particle.
Taking the determinant of the second of the equations (\ref{4.6.1}) gives $\gamma=0$,
so that the lagrangian (\ref{2.3.1}) vanishes,  this does not necessarily mean that
variations in the action vanish;  in the particular case of (\ref{2.2.1}) the action
describes null geodesics or in other word the particle is massless.
Substituting in the momentum constraint \ref{2.3.c} and dividing by $-\hbar$ gives
\be
v^av^b\bx\ps+v^av^{b\mu}\ps_\mu-\fr{k^2}{\hbar^2}\ga\ga^{ab}\ps=0,
\label{4.6.2}
\ee
using the second part of (\ref{4.6.1}) and assuming that $v^{b\mu}\ps_\mu=0$
this is just the Klein-Gordon equation with $k^2=m^2$.
Applying (\ref{4.6.1}) to the extended momentum constraint (\ref{2.3.b})
and dividing by $-i\hbar$ gives
\be
P^\mu_x\ps^a_\mu+k{\rm p}\sq{-\ga}r^{\mu a}\ps_{\mu a}=0,
\label{4.6.b}
\ee
which reduces to (\ref{4.1.b}) when ${\rm p}=0$,
the same assumptions about operator ordering have been made.
Applying (\ref{4.6.1}) to the relative motion equation (\ref{2.3.5})
making similar assumptions as those for the point particle
and dividing by $-k(p+1)$ gives
\be
\p_a\left[\sq{-\ga}r^a_\bt\left(g^{\al\bt}-\fr{\hbar^2}{k^2}\na^\al\na^\bt\right)\ps\right]
=\fr{\hbar^2}{k^2}\sq{-\ga}R^\al_{.\de\ga\bt}r^\bt\na^\ga\na^\de\ps.
\label{4.6.3}
\ee
The form of this equation allows the question:
''Did the Universe begin with a quantum fluctuation?'' to be addressed \cite{tyron}.
One must have initially that the Riemann tensor vanishes otherwise the branes are moving
with respect to one another via the classical equations of relative motion (\ref{2.3.5}),
also lack of movement implies $r^{\bt a}=0$,  so that the equation (\ref{4.6.3})
is of the form $0\times f(\ps)=0$,
implying that no fluctuation of $\ps$ can produce a beginning.

Alternatively one can substitute for the extended momentum
\be
P^{\mu a}_x\rightarrow-i\hbar v^a\na^\mu.
\label{4.6.4}
\ee
The constraints between the momentum $p$ remain unaffected.
The extended constraint (\ref{2.3.b}) becomes
\be
-i\hbar p_{\mu a}v^a\ps^\mu+k{\rm p}\sq{-\ga}r^{\bt a}p_{\bt a}=0.
\label{4.6.e}
\ee
The relative motion equation (\ref{2.3.5}) becomes
\be
\p_a\left[v^a\ps^\mu\right]
=\fr{-i}{k\hbar\sq{-\ga}}R^\mu_{.\de\ga\bt}r^\bt p^{\ga a}p^\de_a\ps.
\label{4.6.5}
\ee
Initially,  as before one can take the rhs=0,
then taking $v^a\ps^\mu_a>v^a_{.a}\ps^\mu$,  $a=\ta$ only and $v^\ta\approx{\rm constant}$
gives $\ps^{\mu\ta}=k$,  suggesting that $\ps$ can linearly increase.
Linear increase is not really a fluctuation,
as for that one would expect trigonometric or harmonic functions,
but it does suggest quantum fluctuations between branes is more likely than in a brane.
\section{Conclusion.}
\label{conc}
The lagrangian formulation of geodesic deviation is readily generalized to describe the
relative motion of other systems,  in particular membranes.
The resulting systems are cross-conjugate rather than self-conjugate,
and this causes problems when approximating or looking for separation of variables solutions.
The corresponding quantum equations are difficult to find,
here some quantum equations which give good classical limits are presented.
These equations allow us to address the question of whether the Universe began with a
quantum fluctuation.   What ''begin'',  ''time'' or why the Universe should be in a state
describable by branes is not looked at,  what is looked at is merely whether the equations
allow one to go from a static flat state to a dynamical one.
If the initial state is not taken to be static then there is a priori state
so that the state is not initial.
It is found that this is not possible for a change in wavefunction inside a brane,
but is for a change in wavefunction between branes.
The change in wavefunction is not initially periodic,
although it might become so once the rhs of (\ref{4.6.5}) becomes non-vanishing.
Thus the change might not be a fluctuation,
but rather just an increase initially of approximately linear form (a twitch);
so that it looks as if the origin of the Universe was more likely caused
by a twitch between branes rather than a twitch in the brane.
\section{Acknowledgement.}
I would like to tank the referees at cejp for pointing out that (\ref{4.6.1})
implies $\gamma=0$.

\end{document}